\documentclass[12pt]{article}   
\newcommand{\mysection}{\setcounter{equation}{0}\section}

\def\beq{\begin{equation}}
\def\eeq{\end{equation}}
\def\beqa{\begin{eqnarray}}
\def\eeqa{\end{eqnarray}}

\newlength{\dinwidth} \newlength{\dinmargin}
\begin{document}
\begin {flushright}
Cavendish-HEP-03/02\\
\end {flushright} 
\vspace{3mm}
\begin{center}
{\Large \bf A unified approach to NNLO soft and virtual corrections
in electroweak, Higgs, QCD, and SUSY processes}
\end{center}
\vspace{2mm}
\begin{center}
{\large Nikolaos Kidonakis}\\
\vspace{2mm}
{\it Cavendish Laboratory\\ University of Cambridge\\
Madingley Road\\ Cambridge CB3 0HE, UK}
\end{center}

\begin{abstract}
I present a unified approach to calculating the next-to-next-to-leading order
(NNLO) soft and virtual QCD corrections to cross sections for 
electroweak, Higgs, QCD, and SUSY processes. 
I derive master formulas that can be used for any 
of these processes in hadron-hadron and lepton-hadron collisions. 
The formulas are based on a unified threshold resummation formalism 
and can be applied to both total and differential cross sections for 
processes with either simple or complex color flows and for various 
factorization schemes and kinematics. 
As a test of the formalism, I rederive known NNLO results for Drell-Yan 
and Higgs production, deep inelastic scattering, and $W^+ \gamma$ production,
and I obtain expressions for several two-loop anomalous dimensions and other
quantities needed in next-to-next-to-leading-logarithm (NNLL) resummations. 
I also present new results for the production of supersymmetric 
charged Higgs bosons;  massive electroweak vector bosons; photons;
heavy quarks in lepton-hadron and hadron-hadron collisions and in 
flavor-changing neutral current processes; jets; and squarks and gluinos. 
The NNLO soft and virtual corrections are often dominant, especially
near threshold. Thus, a unified approach to these corrections is important 
in the search for new physics at present and future colliders.
\end{abstract}

\thispagestyle{empty} \newpage \setcounter{page}{2}

\mysection{Introduction}

Progress in theoretical particle physics, from electroweak theory
and Quantum Chromodynamics in the Standard Model to supersymmetry and beyond,
often involves the comparison of predictions of theory with 
data from high-energy hadron-hadron and lepton-hadron collisions.
Recent theoretical advances include an array of resummations
and a few next-to-next-to-leading order (NNLO) calculations 
\cite{WGQCD,QCDSM}. These advances are necessary in many cases
where next-to-leading order (NLO) calculations are not accurate enough,
since higher-order corrections reduce the scale dependence and increase 
theoretical accuracy.

Next-to-next-to-leading order calculations are technically very
challenging and have been completed only for a few processes,
including Drell-Yan \cite{DY1,DY2} and Higgs production 
\cite{Higgs1,Higgs2,Higgs3}
and deep inelastic scattering \cite{DIS1,DIS2}.
The corrections are usually split into hard, soft,
and virtual parts, corresponding to contributions from energetic, soft, 
and virtual gluons, respectively.
The soft and virtual corrections are an important component of  
the total result both theoretically and numerically. In fact, in some
schemes and kinematical regions, e.g. threshold, they are the dominant part.
Highlighting their importance is the fact that
for Drell-Yan \cite{DY3} and Higgs \cite{Higgs4,Higgs5} 
production the NNLO soft and virtual corrections were presented before 
the full NNLO result was calculated.

We will see that there is a universality in the form of these
corrections, which becomes more evident from the techniques of threshold 
resummations, which arise from factorization properties of the cross sections.
Threshold corrections can be resummed to all orders, and finite-order
expansions of resummed cross sections have provided us with many cross sections
at NNLO and next-to-next-to-leading logarithm (NNLL) accuracy \cite{NKrev}.
Thus it is a worthwhile aim to see if a unified approach can be 
given for the calculation of NNLO total and differential cross sections
for any process in the Standard Model
and beyond in hadron-hadron and hadron-lepton colliders and in various
factorization schemes and kinematics. This will not only increase 
and deepen theoretical understanding but will also help avoid 
effort duplication in calculations 
of NNLO corrections for new processes. It is important to note that new
particles, such as in supersymmetry, will likely be discovered
near threshold where the soft and virtual corrections are important.
The new unified approach to NNLO soft and virtual corrections is the 
topic of this paper.

The calculation of cross sections in hadron-hadron or lepton-hadron
collisions can be written schematically as
\beqa
\sigma=\sum_f \int \left[ \prod_i  dx_i \, \phi_{f/h_i}(x_i,\mu_F^2)\right]\,
{\hat \sigma}(s,t_i,\mu_F,\mu_R) \, ,
\label{factcs}
\eeqa
where $\sigma$ is the physical cross section,
$\phi_{f/h_i}$ is the distribution function for parton $f$ carrying momentum 
fraction $x_i$ of hadron $h_i$, at a factorization scale $\mu_F$, 
while $\mu_R$ is the renormalization scale.
The parton-level hard scattering cross section is denoted by 
${\hat \sigma}$, and $s$ and $t_i$ are standard kinematical invariants. 
In a lepton-hadron collision we obviously have one parton distribution
($i=1$) while in a hadron-hadron collision $i=1,2$.
We note here that $\sigma$ and ${\hat \sigma}$ are not restricted to be
total cross sections; they can represent any relevant differential cross 
section of interest.

In general, $\hat{\sigma}$ includes plus distributions ${\cal D}_l(x_{th})$ 
and delta functions $\delta(x_{th})$ with respect to a kinematical variable 
$x_{th}$ that measures distance from threshold, with $l \le 2n-1$ at $n$th 
order in $\alpha_s$ beyond the leading order. These are the soft and 
virtual corrections. In single-particle inclusive (1PI) kinematics, 
$x_{th}$ is usually called $s_4$
(or $s_2$), $s_4=s+t+u-\sum m^2$, and vanishes at threshold. Then
\beq
{\cal D}_l(s_4)\equiv\left[\frac{\ln^l(s_4/M^2)}{s_4}\right]_+
=\frac{\ln^l(s_4/M^2)}{s_4} \theta(s_4-\Delta)
+\frac{1}{l+1} \ln^{l+1}\left(\frac{\Delta}{M^2}\right) \delta(s_4) \, ,
\eeq
where $\Delta$ is a small parameter introduced in order to separate
the hard $x_{th} > \Delta$ and soft $x_{th} < \Delta$ gluon regions,
and $M^2$ is a hard scale relevant to the process at hand, for example
the mass $m$ of a heavy quark, the transverse momentum $p_T$ of a jet, etc.
In pair-invariant-mass (PIM) kinematics, with $Q^2$ the invariant mass 
squared of the produced pair, $x_{th}$ is usually called $1-x$ or $1-z$, 
with $z=Q^2/s \rightarrow 1$ at threshold.  Then
\beq
{\cal D}_l(z)\equiv\left[\frac{\ln^l(1-z)}{1-z}\right]_+
=\frac{\ln^l(1-z)}{1-z} \theta(1-z-\Delta)
+\frac{1}{l+1} \ln^{l+1}(\Delta) \, \delta(1-z) \, .
\eeq
The highest powers of these distributions at each order in $\alpha_s$
are the leading logarithms (LL),
the second highest are the next-to-leading logarithms (NLL), 
the third highest are the next-to-next-to-leading logarithms (NNLL), 
etc. These logarithms can be resummed to all orders in perturbation theory.
By now there are several processes for which NLL resummations and
NNLO-NNLL results (i.e. the NNLL terms at NNLO) have been presented 
\cite{NKrev}.

In this paper I will present master formulas for the NLO and NNLO 
soft and virtual corrections for any process in hadron-hadron and
lepton-hadron collisions.
In the next section, I present a threshold resummation formula,
that builds on and unifies previous work \cite{NKrev,CLS,KS,KOS,LOS,NK}. 
I then present master formulas for the NLO and NNLO soft and virtual 
corrections in processes with simple color flow, 
that arise from the expansion of the resummation
formula and matching to NLO. The formulas cover both the $\overline{\rm MS}$
and DIS schemes, and 1PI and PIM kinematics. In Section 3, I present 
results for electroweak/Higgs processes as well as QCD and SUSY processes 
with simple color flows.
I rederive known NNLO results for the Drell-Yan process and
Higgs production, as well as for deep inelastic scattering
and $W^+ \gamma$ production, thus obtaing expressions for two-loop
anomalous dimensions and other quantities that are universal in 
quark-antiquark and gluon-gluon scattering and are needed for NNLL 
resummations. I also present new results for the NNLO corrections 
for supersymmetric charged Higgs production, $W,Z$ plus jet production, 
direct photon production,
DIS heavy quark production, as well as single-top quark production  mediated
by flavor-changing neutral currents.
In Section 4, I extend the master formulas to processes with complex color 
flows and I present results for heavy quark hadroproduction, 
jet production, and squark and gluino production.
I close with a discussion of extensions to higher orders and conclusions 
in Section 5. 

\mysection{NNLO master formula for soft and virtual corrections}

\subsection{Soft corrections from threshold resummation}

In hadron-hadron collisions we study processes where partons
$f_i$ collide and produce a specific final state. 
The partonic processes are of the form
\beq
f_{1}(p_1)\, + \, f_{2}(p_2) \rightarrow F\, + \, X \, ,
\eeq
where $F$ represents a system in the final state, and $X$ any additional
allowed final-state particles.
So $F$ can represent a pair of heavy quarks, or a single heavy quark,
or jet, or photon, a Higgs boson, squarks, etc.

In lepton-hadron collisions the processes are of the form
\beq
f_{1}(p_1) \, + \, l(p_2) \rightarrow F\, + \, X \, .
\eeq
In either case $s=(p_1+p_2)^2$ and $t_i$ are the usual Mandelstam invariants
formed by the four-momenta of the particles in the scattering.

Resummed cross sections have by now been studied for a variety of processes
\cite{NKrev}.
The resummation of threshold logarithms is carried out in moment space.
We define moments of the partonic cross section by
${\hat\sigma}(N)=\int dz \, z^{N-1} {\hat\sigma}(z)$ or by  
${\hat\sigma}(N)=\int (ds_4/s) \;  e^{-N s_4/s} {\hat\sigma}(s_4)$,
with $N$ the moment variable.
The resummed partonic cross section in moment space is then given by 
\beqa
{\hat{\sigma}}^{res}(N) &=&   
\exp\left[ \sum_i E^{(f_i)}(N_i)\right] \; 
\exp\left[ \sum_j {E'}^{(f_j)}(N_j)\right] 
\nonumber\\ && \hspace{-10mm} \times \,
\exp \left[\sum_i 2\int_{\mu_F}^{\sqrt{s}} {d\mu' \over \mu'}
\left(\frac{\alpha_s(\mu'^2)}{\pi}\gamma_i^{(1)}
+{\gamma'}_{i/i}\left(\alpha_s(\mu'^2)\right)\right)\right] \;
\exp\left[2\, d_{\alpha_s} \int_{\mu_R}^{\sqrt{s}}\frac{d\mu'}{\mu'} 
\beta\left(\alpha_s(\mu'^2)\right)\right] 
\nonumber\\ && \hspace{-10mm} \times \,
{\rm Tr} \left \{H\left(\alpha_s(\mu_R^2)\right) \;
\bar{P} \exp \left[\int_{\sqrt{s}}^{{\sqrt{s}}/{\tilde N_j}} 
{d\mu' \over \mu'} \;
{\Gamma'}_S^\dagger\left(\alpha_s(\mu'^2)\right)\right] \;
{\tilde S} \left(\alpha_s(s/{\tilde N_j}^2) \right) \right. 
\nonumber\\ && \quad \left.\times \,
P \exp \left[\int_{\sqrt{s}}^{{\sqrt{s}}/{\tilde N_j}} 
{d\mu' \over \mu'}\; {\Gamma'}_S
\left(\alpha_s(\mu'^2)\right)\right] \right\} \, .
\nonumber \\ 
\label{resHS}
\eeqa
The sums over $i$ run over incoming partons: in hadron-hadron
colisions we have two partons in the initial state, so $i=1,2$; 
in lepton-hadron collisions
we only have one parton (and one lepton). The sum over $j$ is relevant
only if we have massless partons in the final state at lowest order. 
Clearly the second exponent is then absent for proceses such as Drell-Yan, 
Higgs, and top quark pair production.
  
Equation (\ref{resHS}) is actually valid for both 1PI and PIM kinematics
with appropriate definitions for $N_i$ and $N_j$.
In 1PI kinematics $N_i=N (-t_i/M^2)$ for incoming partons $i$,
and  $N_j=N (s/M^2)$ for outgoing massless partons $j$; here $M^2$ is any
chosen hard scale relevant to the process at hand.
The kinematical invariants $t_i$ are assigned through
$S_4/S=s_4/s-\sum_i(1-x_i)t_{i}/s$ \cite{LOS,KV} with $S_4$, $S$, the hadronic
analogs of $s_4$, $s$; note that the $t_i$ may include a $-m^2$ in case
of massive particles. 
In PIM kinematics $N_i=N_j=N$. Often the resummed cross sections are
given with the choice $M^2=s$ in which case $N_j=N$ even in 1PI kinematics; 
here we keep our expressions more general.
Also note that ${\tilde N}=N e^{\gamma_E}$, with $\gamma_E$ the Euler constant.

The first exponent in Eq. (\ref{resHS}) resums the $N$-dependence of
incoming partons \cite{GS,CT} and is given in the $\overline{\rm MS}$ scheme 
by \cite{NK}
\beq
E^{(f_i)}(N_i)=
-\int^1_0 dz \frac{z^{N_i-1}-1}{1-z}\;
\left \{\int^{\mu_F^2}_{(1-z)^2s} \frac{d\mu'^2}{\mu'^2}
A^{(f_i)}\left(\alpha_s({\mu'}^2)\right)
+{\nu}_{f_i}\left[\alpha_s((1-z)^2 s)\right]\right\} \, ,
\label{Eexp}
\eeq
with $A^{(f_i)}(\alpha_s) = C_{f_i} [ {\alpha_s/\pi}+({\alpha_s/\pi})^2 K/2]
+\cdots$.
Here $C_{f_i}=C_F=(N_c^2-1)/(2N_c)$ for an incoming quark or antiquark,
and $C_{f_i}=C_A=N_c$ for an incoming gluon, with $N_c$ the number of colors,
while $K= C_A\; ( 67/18-\pi^2/6 ) - 5n_f/9$,
where $n_f$ is the number of quark flavors.
Also ${\nu}_{f_i}=(\alpha_s/\pi)C_{f_i}+(\alpha_s/\pi)^2 {\nu}_{f_i}^{(2)}
+\cdots$. 

The second exponent resums the $N$-dependence of any outgoing massless
partons and is given by \cite{LOS,KV}
\beq
{E'}^{(f_j)}(N_j)=
\int^1_0 dz \frac{z^{N_j-1}-1}{1-z}\;
\left \{\int^{1-z}_{(1-z)^2} \frac{d\lambda}{\lambda}
A^{(f_j)}\left(\alpha_s\left(\lambda s\right)\right)
-B'_j\left[\alpha_s((1-z)s)\right]
-\nu_{j}\left[\alpha_s((1-z)^2 s)\right]\right\} \, ,
\label{Ejexp}
\eeq
where $B'_j=(\alpha_s/\pi){B'}_j^{(1)}+(\alpha_s/\pi)^2 {B'}_j^{(2)}+\cdots$
with ${B'}_q^{(1)}=3C_F/4$ and ${B'}_g^{(1)}=\beta_0/4$.
We will determine ${B'}_q^{(2)}$ in Section 3.

The $\gamma_i^{(1)}$ in the third exponent are one-loop parton anomalous 
dimensions:  $\gamma_q^{(1)}=3C_F/4$ and $\gamma_g^{(1)}=\beta_0/4$ 
for quarks and gluons, respectively; it is important to note that in 
this specific form that the resummed cross section has been written, 
with $\mu_F^2$ as the upper limit of the integral over $d{\mu'}^2$ in 
$E^{(f_i)}(N_i)$, only the one-loop $\gamma_i$ appears in the third exponent. 
We also have defined ${\gamma'}_{i/i}$ as the moment-space 
anomalous dimension of the ${\overline {\rm MS}}$ density $\phi_{i/i}$,
minus it's one-loop and its $N$-dependent two-loop components. 
This is again due to the specific form that we use for the resummed cross 
section. Thus, ${\gamma'}_{i/i}^{(2)}$ is the $N$-independent part
of the two-loop anomalous dimension ${\gamma}_{i/i}$ \cite{GALY,GFP}
and is given for quarks and gluons by 
\beq
{\gamma'}_{q/q}^{(2)}=C_F^2\left(\frac{3}{32}-\frac{3}{4}\zeta_2
+\frac{3}{2}\zeta_3\right)
+C_F C_A\left(-\frac{3}{4}\zeta_3+\frac{11}{12}\zeta_2+\frac{17}{96}\right)
+n_f C_F \left(-\frac{\zeta_2}{6}-\frac{1}{48}\right)\, ,
\eeq
and
\beq
{\gamma'}_{g/g}^{(2)}=C_A^2\left(\frac{2}{3}+\frac{3}{4}\zeta_3\right)
-n_f\left(\frac{C_F}{8}+\frac{C_A}{6}\right) \, , 
\eeq
respectively, with $\zeta_2=\pi^2/6$ and $\zeta_3=1.2020569\cdots$.

The $\beta$ function in the fourth exponent is given by
$\beta(\alpha_s) \equiv \mu \, d \ln g/d \mu
=-\beta_0 \alpha_s/(4 \pi)-\beta_1 \alpha_s^2/(4 \pi)^2 +\cdots$, 
with $\beta_0=(11C_A-2n_f)/3$ and $\beta_1=34 C_A^2/3-2n_f(C_F+5C_A/3)$.
The constant $d_{\alpha_s}=0,1,2$ if the Born cross section is of order
$\alpha_s^0$, $\alpha_s^1$, $\alpha_s^2$, respectively.

The trace appearing in the resummed expression is taken in the space
of color exchanges.
The symbols $P$ and ${\bar P}$ denote path ordering in the same sense
as the variable $\mu'$ and against it, respectively.
The evolution of the soft function from scale ${\sqrt s}/{\tilde N}_j$ to 
$\sqrt s$ follows from its renormalization group properties and 
is given in terms of the soft anomalous dimension matrix $\Gamma_S$
\cite{KS,KOS,NKrev}. 
In Eq. (\ref{resHS}) we actually use ${\Gamma'}_S$, which is given by
$\Gamma_S$ after dropping all gauge-dependent terms. We can do that 
because the gauge dependence has been shown to cancel out.
At one loop, the gauge terms are of the form $C_{f_i} \ln(2 \nu^n_i)$, where
$\nu^n_i=(v_i \cdot n)^2/|n|^2$ with  $v_i$ a velocity vector and $n$ 
the axial gauge vector \cite{NKrev,KS,KOS}. 
In processes with simple color flow, $\Gamma_S$ is a trivial $1\times 1$ 
matrix. For the determination of $\Gamma_S$ in processes with complex 
color flow, an appropriate choice of color basis has to be made.
For gluon-gluon scattering, the most complex color flow encountered,
$\Gamma_S$ is an $8\times 8$ matrix \cite{KOS}.
The process-dependent soft anomalous dimension matrices, evaluated through 
the calculation of eikonal vertex corrections \cite{KS}, have by now been 
presented at one loop for practically all partonic processes \cite{NKrev}. 
They can be explicitly calculated for any process using the techniques and 
results in Refs. \cite{KS,NKrev}.
Work is currently being done on 
two-loop calculations of these anomalous dimensions \cite{NK2l}, 
but we note that we can extract the universal component of these
anomalous dimensions for quark-antiquark and gluon-gluon
initiated processes from the Drell-Yan and Higgs NNLO results as detailed
in Section 3. 
In the color bases that we normally use, the soft matrices, $S$, are diagonal.
At lowest order, the trace of the product of the hard and soft matrices
reproduces the Born cross section for each partonic process.
We also note that the $\Gamma_S$ matrices are in general not diagonal in 
the color bases that we use for complex color flows. 
If we perform a diagonalization so that
the $\Gamma_S$ matrices do become diagonal, then the path-ordered
exponentials of matrices in the resummed expression
reduce to simple exponentials; however, this diagonalization
procedure is complicated in practice \cite{NKrev}.
A finite-order expansion bypasses the need for this 
diagonalization procedure.
For the discussion below we expand ${\Gamma'}_S$ as
${\Gamma'}_S=(\alpha_s/\pi){\Gamma'}_S^{(1)}
+(\alpha_s/\pi)^2{\Gamma'}_S^{(2)}+\cdots$.
We will determine the universal component of ${\rm Re}{\Gamma'}_S^{(2)}
-\nu^{(2)}$, where ``Re'' stands for ``real part of,''  for quark-antiquark 
and gluon-gluon processes in Section 3.

Finally, we note that there are different ways of writing threshold
resummation formulas that have been presented in the past for various
processes, all consistent or equivalent at NLL accuracy; the expression 
presented here unifies those expressions for arbitrary processes and 
is superior in its simplicity and generality.

In the DIS scheme, the resummed cross section may be written as
\beqa
{\hat{\sigma}}^{res}_{\rm DIS}(N) &=&
{\hat{\sigma}}^{res}_{\overline{\rm MS}}(N) 
\exp\left\{-\sum_i \int^1_0 dz \frac{z^{N_i-1}-1}{1-z}\;
\left[\int^{1-z}_{1} \frac{d\lambda}{\lambda}
A^{(f_i)}\left(\alpha_s(\lambda \mu_F^2)\right)
-B_{f_i}\left[\alpha_s((1-z) \mu_F^2)\right]\right]\right\} \, ,
\nonumber \\
\label{resDIS}
\eeqa
where ${\hat{\sigma}}^{res}_{\overline{\rm MS}}$ is the 
${\overline{\rm MS}}$ cross section in Eq. (\ref{resHS}), and
$B_{f_i}=(\alpha_s/\pi)B_{f_i}^{(1)}+(\alpha_s/\pi)^2 B_{f_i}^{(2)}
+\cdots$.
For quarks, to which the scheme is usually applied, $B_q^{(1)}=3C_F/4$.
We will determine $B_q^{(2)}$ in Section 3.

We first expand the resummed formulas in Eqs. (\ref{resHS})
and (\ref{resDIS})  for processes
with simple color flow, i.e. when $H$, $S$ and $\Gamma_S$ are
$1\times 1$ matrices, to next-to-leading
order and present NLO master formulas in the $\overline{\rm MS}$ 
and DIS schemes, which reproduce the NLO results
for a variety of processes. We then present NNLO master formulas for
soft and virtual corrections in processes with simple color flow
in both the $\overline{\rm MS}$ and DIS schemes.

\subsection{NLO master formula for soft and virtual corrections 
- simple color flow}

\subsubsection{$\overline{\rm MS}$ scheme}

At next-to-leading order, the expansion of Eq. (\ref{resHS}) gives
the NLO soft and virtual corrections in the $\overline{\rm MS}$ scheme:
\beq
{\hat{\sigma}}^{(1)} = \sigma^B \frac{\alpha_s(\mu_R^2)}{\pi}
\left\{c_3 \, {\cal D}_1(x_{th}) + c_2 \, {\cal D}_0(x_{th}) 
+c_1 \, \delta(x_{th})\right\} \, ,
\label{NLOMS}
\eeq
where $\sigma^B$ is the Born term,
\beq
c_3=\sum_i 2C_{f_i} -\sum_j C_{f_j}\, ,
\label{c3}
\eeq
and
\beq
c_2=2 \, {\rm Re}{\Gamma'}_S^{(1)}- \sum_i \left[C_{f_i}
+2 C_{f_i} \, \delta_K \, \ln\left(\frac{-t_i}{M^2}\right)+
C_{f_i} \ln\left(\frac{\mu_F^2}{s}\right)\right]
-\sum_j \left[{B'}_j^{(1)}+C_{f_j}
+C_{f_j}\, \delta_K \, \ln\left(\frac{M^2}{s}\right)\right] \, ,
\eeq
where ${\rm Re}{\Gamma'}_S^{(1)}$ is the real part of the one-loop 
$\Gamma_S$ after dropping all gauge-dependent terms, and $\delta_K$ is
$0$ for PIM kinematics and $1$ otherwise. We remind the reader
that the sums over $i$ run over incoming partons and the sums over $j$
run over any massless partons in the final state at lowest order. 
For future use we will write $c_2=c_2^{\mu}+T_2$, where
$c_2^{\mu}$ represents the scale term $-\sum_i C_{f_i} \ln(\mu_F^2/s)$
and $T_2$ is the remainder.
Also,
$c_1 =c_1^{\mu} +T_1$, with
\beq
c_1^{\mu}=\sum_i \left[C_{f_i}\, \delta_K \, \ln\left(\frac{-t_i}{M^2}\right) 
-\gamma_i^{(1)}\right]\ln\left(\frac{\mu_F^2}{s}\right)
+d_{\alpha_s} \frac{\beta_0}{4} \ln\left(\frac{\mu_R^2}{s}\right) \, .
\label{c1mu}
\eeq
$T_1$ denotes the terms in $c_1$ that do not involve the factorization and 
renormalization scales,  and it can be read off 
from a full calculation of the NLO virtual corrections for any 
specified process; a formal expression for these terms is given in Section 
4.1 in the more general case of complex color flow.

We note that our formula passes a number of tests. As we will see, 
its predictions agree with exact NLO soft plus virtual results for all 
processes where those results are already available. Also the renormalization 
and factorization scale dependence in the physical cross section 
(after convoluting the partonic cross section with the parton distributions)
cancels explicitly, i.e. $d\sigma/d\mu_F=0$ and
$d\sigma/d\mu_R=0$ at NLO.

\subsubsection{DIS scheme}

In the DIS scheme, the NLO soft and virtual corrections are
\beq
{\hat{\sigma}}^{(1)}_{\rm DIS} = \sigma^B \frac{\alpha_s(\mu_R^2)}{\pi}
\left\{c_3' \, {\cal D}_1(x_{th}) + c_2' \, {\cal D}_0(x_{th}) 
+c_1' \, \delta(x_{th})\right\} \, ,
\label{NLODIS}
\eeq
with
$c_3'=c_3-\sum_i C_{f_i}$, 
\beq
c_2'=c_2+\sum_i \left[C_{f_i} \, \delta_K \, 
\ln\left(\frac{-t_i}{M^2}\right)+B_{f_i}^{(1)}
\right] \, ,
\eeq
and
\beq
c_1'=c_1-\sum_i \left[C_{f_i} \, \delta_K \,
\frac{1}{2}\ln^2\left(\frac{-t_i}{M^2}\right)
+B_{f_i}^{(1)} \, \delta_K \, 
\ln\left(\frac{-t_i}{M^2}\right)-D_{f_i}\right]\, ,
\eeq
with $c_3$, $c_2$, $c_1$ the $\overline{\rm MS}$ results given in the 
previous subsection. For future use we will write $c'_2={c'_2}^{\mu}+T'_2$, 
and $c'_1={c'_1}^{\mu}+T'_1$ as we did for the $\overline{\rm MS}$ corrections.
Note that the changes in going from the $\overline{\rm MS}$ scheme
to the DIS scheme are all in the scale-independent
parts of the $c_i$'s, i.e. $c_2^{\mu}$ and $c_1^{\mu}$ remain unchanged
while $T_2$ and $T_1$ (and $c_3$) are modified when changing schemes.
The DIS scheme is normally applied to quarks. For quarks the term
$D_{f_i}$ in $c_1'$ is $D_q=C_F \zeta_2+(9/4)C_F$.  
We note that our formula passes the same tests as we outlined in the
previous subsection for the $\overline{\rm MS}$ scheme. 

\subsection{NNLO master formula for soft and virtual corrections 
- simple color flow}

\subsubsection{$\overline{\rm MS}$ scheme}

At next-to-next-to-leading order, the expansion of Eq. (\ref{resHS}), with 
matching to the full NLO soft-plus-virtual result, gives the NNLO
soft and virtual corrections in the $\overline{\rm MS}$ scheme
\beq
{\hat\sigma}^{(2)}=
\sigma^B \frac{\alpha_s^2(\mu_R^2)}{\pi^2} \, {\hat{\sigma'}}^{(2)}
\label{NNLOm}
\eeq
with
\beqa
{\hat{\sigma'}}^{(2)}&=& 
\frac{1}{2} c_3^2 \, {\cal D}_3(x_{th})
+\left[\frac{3}{2} c_3 \, c_2 
- \frac{\beta_0}{4} c_3
+\sum_j C_{f_j} \frac{\beta_0}{8}\right] {\cal D}_2(x_{th})
\nonumber \\ && \hspace{-10mm}
{}+\left\{c_3 \, c_1 +c_2^2
-\zeta_2 \, c_3^2 -\frac{\beta_0}{2} \, T_2 
+\frac{\beta_0}{4} c_3  \ln\left(\frac{\mu_R^2}{s}\right)
+\sum_i C_{f_i} \, K \right.
\nonumber \\ && \hspace{-10mm} \quad \quad \left.
{}+\sum_j C_{f_j} \left[-\frac{K}{2} 
+\frac{\beta_0}{4} \, \delta_K \, \ln\left(\frac{M^2}{s}\right)\right]
-\sum_j\frac{\beta_0}{4} {B'}_j^{(1)} \right\}
{\cal D}_1(x_{th})
\nonumber \\ && \hspace{-10mm} 
{}+\left\{c_2 \, c_1 -\zeta_2 \, c_2 \, c_3+\zeta_3 \, c_3^2 
-\frac{\beta_0}{2} T_1
+\frac{\beta_0}{4}\, c_2 \ln\left(\frac{\mu_R^2}{s}\right) 
+2 \, {\rm Re}{\Gamma'}_S^{(2)}-\sum_i {\nu}_{f_i}^{(2)}
\right. 
\nonumber \\ && \hspace{-10mm} \quad \quad
{}+\sum_i C_{f_i} \left[\frac{\beta_0}{8} 
\ln^2\left(\frac{\mu_F^2}{s}\right)
-\frac{K}{2}\ln\left(\frac{\mu_F^2}{s}\right)
-K \, \delta_K \, \ln\left(\frac{-t_i}{M^2}\right)\right]
-\sum_j \left({B'}_j^{(2)}+\nu_j^{(2)}\right)
\nonumber \\ && \hspace{-10mm} \quad \quad \left.
{}+\sum_j C_{f_j} \, \delta_K \, \left[\frac{\beta_0}{8}
\ln^2\left(\frac{M^2}{s}\right)
-\frac{K}{2}\ln\left(\frac{M^2}{s}\right)\right]
-\sum_j \frac{\beta_0}{4}
{B'}_j^{(1)} \, \delta_K \, \ln\left(\frac{M^2}{s}\right) \right\}
{\cal D}_0(x_{th})  
\nonumber \\ &&  \hspace{-10mm}
{}+\left\{\frac{1}{2} c_1^2
-\frac{\zeta_2}{2}\, c_2^2
+\frac{1}{4}\zeta_2^2 \, c_3^2
+\zeta_3 \, c_3 \, c_2
-\frac{3}{4}\zeta_4 \, c_3^2 
+\frac{\beta_0}{4} c_1 \ln\left(\frac{\mu_R^2}{s}\right)
+2 \, {\rm Re}{\Gamma'}_S^{(2)} \, \delta_K \, 
\ln\left(\frac{M^2}{s}\right)
\right.
\nonumber \\ && \hspace{-10mm} \quad \quad
{}-\frac{\beta_0}{2} \, \delta_K \, T_1 \ln\left(\frac{M^2}{s}\right)
+\frac{\beta_0}{4} \, \delta_K \, T_2 \ln^2\left(\frac{M^2}{s}\right)
+\frac{d_{\alpha_s}}{16}\left[-\frac{\beta_0^2}{2}
\ln^2\left(\frac{\mu_R^2}{s}\right)
+\beta_1\ln\left(\frac{\mu_R^2}{s}\right)\right]
\nonumber \\ && \hspace{-10mm} \quad \quad 
{}+\sum_i \frac{\beta_0}{8} \left[\gamma_i^{(1)}
-C_{f_i} \, \delta_K \, \ln\left(\frac{-t_i}{M^2}\right)\right] 
\ln^2\left(\frac{\mu_F^2}{s}\right)
+\sum_i C_{f_i} \frac{K}{2} \, \delta_K \, \ln\left(\frac{-t_i}{M^2}\right)
\ln\left(\frac{\mu_F^2}{s}\right)
\nonumber \\ && \hspace{-10mm} \quad \quad 
{}-\sum_i {\gamma'}_{i/i}^{(2)} \ln\left(\frac{\mu_F^2}{s}\right)
+\sum_i C_{f_i}  \, \delta_K \, \left[\frac{\beta_0}{6}
\ln^3\left(\frac{-t_i}{M^2}\right)
+\left(\frac{\beta_0}{4}+\frac{K}{2}\right)
\ln^2\left(\frac{-t_i}{M^2}\right)\right]
\nonumber \\ && \hspace{-10mm} \quad \quad 
{}+\sum_i C_{f_i} \frac{\beta_0}{2} \, \delta_K \,\ln\left(\frac{M^2}{s}\right)
\left[\ln^2\left(\frac{-t_i}{M^2}\right)-\ln\left(\frac{-t_i}{M^2}\right)
\ln\left(\frac{M^2}{s}\right)-\frac{1}{2}\ln\left(\frac{M^2}{s}\right)
+\ln\left(\frac{-t_i}{M^2}\right)\right]
\nonumber \\ && \hspace{-10mm} \quad \quad 
{}+\sum_i {\nu}_{f_i}^{(2)}  \, \delta_K \, \ln\left(\frac{-t_i}{M^2}\right)
-\sum_j \left({B'}_{j}^{(2)}+{\nu}_{j}^{(2)}\right)  \, \delta_K \, 
\ln\left(\frac{M^2}{s}\right) 
\nonumber \\ && \hspace{-10mm} \quad \quad \left.
{}+\sum_j \left[\frac{\beta_0}{8} C_{f_j}\ln\left(\frac{M^2}{s}\right)
-\frac{K}{4} C_{f_j}-\frac{\beta_0}{8} {B'}_j^{(1)}
\right]  \, \delta_K \, \ln^2\left(\frac{M^2}{s}\right)
+ R \right\} \delta(x_{th}) \, ,
\label{NNLOmaster}
\eeqa
where the last term $R$, for which a formal expression 
is given in Section 4.1 in the more general case of complex color flow, 
can only be known from a full two-loop calculation. 
We will derive in Section 3 the universal components of $R$ for processes
with quark-antiquark and gluon-gluon collisions by comparing our predictions
to the full NNLO corrections for Drell-Yan and Higgs production. 

The quantities
$d_{\alpha_s}$, $\beta_0$, $\beta_1$, $K$, $\zeta_2$, and $\zeta_3$
have all been defined in Section 2.1, and $\zeta_4=\pi^4/90$.
We note that ${\rm Re}{\Gamma'}_S^{(2)}$ is the real part of the two-loop 
$\Gamma_S$ after dropping all gauge-dependent terms.
The universal part of $2{\rm Re}{\Gamma'}_S^{(2)}-\sum_i\nu_{f_i}^{(2)}$ 
in quark-antiquark and gluon-gluon collisions will be derived in 
Section 3 from Drell-Yan and Higgs production. 
${B'}_q^{(2)}$ will also be derived in Section 3.
There is also current work on 
full two-loop evaluations for general processes \cite{NK2l}. 
Also note, again, that the sum over $i$ involves summing over the
two incoming partons in hadron-hadron collisions, or one parton
in lepton-hadron collisions, and that there is no sum over $j$ if there
are no massless partons in the final state at lowest order.
Also note that $\delta_K$ in the formula again indicates which terms vanish
in PIM kinematics. Finally, we note that the choice $M^2=s$ further
reduces the number of terms in the master formula in 1PI kinematics; 
however, we keep our expression as general as possible.

As we will see below, the NNLO master formula passes many rigorous
tests. It reproduces the exact NNLO soft plus virtual results 
for Drell-Yan and Higgs production, and deep inelastic scattering, 
as well as the NNLO-NNLL results that have been derived already for 
many different processes from threshold resummation studies. 
Also I have checked explicitly that at NNLO the renormalization and
factorization scale dependence in the physical cross section (after 
convoluting the partonic cross section with the parton distributions) 
cancels out for both hadron-hadron and lepton-hadron collisions.

\subsubsection{DIS scheme}

In the DIS scheme the NNLO soft and virtual corrections are
\beq
{\hat\sigma}^{(2)}_{\rm DIS}=
\sigma^B \frac{\alpha_s^2(\mu_R^2)}{\pi^2} \, {\hat{\sigma'}}^{(2)}_{\rm DIS}
\label{NNLOd}
\eeq
with
\beqa
{\hat{\sigma'}}^{(2)}_{\rm DIS}&=& 
{\hat{\sigma'}}^{(2)}|_{c_i'}
-\sum_i \frac{\beta_0}{8} C_{f_i} \, {\cal D}_2(x_{th})
\nonumber \\ && \hspace{-15mm}
{}+\left[\frac{\beta_0}{4}\left(T'_2-T_2\right)
-\sum_i C_{f_i} \frac{K}{2} +\sum_i C_{f_i} \frac{\beta_0}{4} 
\ln\left(\frac{\mu_F^2}{s}\right)\right] {\cal D}_1(x_{th})
\nonumber \\ && \hspace{-15mm}
{}+\left\{\frac{\beta_0}{4}\left(T'_1-T_1\right)
+\sum_i C_{f_i} \, \delta_K \, \frac{K}{2} \ln\left(\frac{-t_i}{M^2}\right)
-\frac{\beta_0}{4} \ln\left(\frac{\mu_F^2}{s}\right)  
\left(T'_2-T_2\right)
+\sum_i B_{f_i}^{(2)} \right\} {\cal D}_0(x_{th})
\nonumber \\ && \hspace{-15mm}
{}+\left\{-\sum_i \frac{\beta_0}{24}C_{f_i} \, \delta_K \,
\ln^3\left(\frac{-t_i}{M^2}\right)
-\sum_i C_{f_i}  \, \delta_K \, \frac{K}{4} \ln^2\left(\frac{-t_i}{M^2}\right)
-\sum_i B_{f_i}^{(1)}  \, \delta_K \, \frac{\beta_0}{8} 
\ln^2\left(\frac{-t_i}{M^2}\right)\right.
\nonumber \\ && \hspace{-15mm} \quad \quad 
{}-\frac{\beta_0}{4} \ln\left(\frac{\mu_F^2}{s}\right)  
\left(T'_1-T_1\right)
+\frac{\beta_0}{2}\, \delta_K \, 
\left(T'_1-T_1\right)\ln\left(\frac{M^2}{s}\right) 
+\sum_i\frac{\beta_0}{12} C_{f_i} \, \delta_K \, 
\ln^3\left(\frac{M^2}{s}\right)
\nonumber \\ && \hspace{-15mm} \quad \quad \left.
{}-\frac{\beta_0}{4}\, \delta_K\, 
\left(T'_2-T_2\right)\ln^2\left(\frac{M^2}{s}\right) 
-\sum_i B_{f_i}^{(2)}  \, \delta_K \,
\ln\left(\frac{-t_i}{M^2}\right) \right\} \delta(x_{th}) \, .
\label{NNLODIS}
\eeqa
We note that the $T'_i$'s are the DIS quantities while the unprimed $T_i$'s
are the $\overline {\rm MS}$ counterparts.
Also ${\hat{\sigma'}}^{(2)}|_{c_i'}$ denotes the 
cross section in Eq. (\ref{NNLOmaster}) after replacing all the $c_i$,
$T_i$, and $R$, by their DIS counterparts $c_i'$, $T_i'$, $R'$.
Our formula reproduces the exact known NNLO corrections for Drell-Yan
and $W^+ \gamma$ production and deep inelastic scattering in the DIS scheme,
as we will see in the next section.
We will derive the expression for $B_q^{(2)}$ in the next section
by matching to the known NNLO corrections for the Drell-Yan process.
We also note that again we have checked the scale independence of the physical
cross section at NNLO.

\mysection{Electroweak/Higgs, QCD, and SUSY processes with 
simple color flows}

We now apply our NLO and NNLO master formulas to a variety of processes
with simple color flow which are of electroweak, QCD, or SUSY origin
at lowest order.

\subsection{The Drell-Yan process}

Our first application is the Drell-Yan process, i.e. lepton pair
production in hadron-hadron collisions, for which
the NNLO corrections have been calculated in
Refs. \cite{DY1, DY2,DY3}. Also a comparison of the expansion of the 
resummed cross section in \cite{GS} with the NNLO corrections
in \cite{DY3} was presented in Ref. \cite{LMag}.
The partonic process we discuss is 
$q {\bar q} \rightarrow V + X$, where $V$ is a vector boson 
($\gamma$, $Z$, $W$) which later decays to a lepton pair 
$V \rightarrow l_1 l_2$ with invariant mass $\sqrt{Q^2}$, 
and $X$ denotes any additional partons in the final state.
At threshold $s=Q^2$, where $s$ is the center-of-mass energy
squared of the incoming quark-antiquark pair. 
The NNLO corrections for the cross section $d\sigma/dQ^2$
in the $\overline{\rm MS}$ scheme are given in Ref. \cite{DY1}.
The soft and virtual corrections are given the label
$\Delta_{q{\bar q}}^{(n),S+V}$ and are given explicitly 
at NLO by Eq. (B.3) in Ref. \cite{DY1}, and at NNLO by Eq. (B.8)
plus the renormalization scale term in Eq. (B.7) of Ref. \cite{DY1}.
Here the plus distributions are $D_l(x)$ with $x=Q^2/s$.

We are able to reproduce these results using our master NLO
and NNLO formulas.
Evidently at lowest order there are no final-state massless partons and 
the cross section is given in PIM kinematics.
We note that for this process ${\rm Re}{\Gamma'}_S^{(1)}=C_F$ and we 
choose the hard scale $M^2=Q^2$.

At NLO our master formula, Eq. (\ref{NLOMS}), gives
\beq
{\hat\sigma}_{q{\bar q} \rightarrow V}^{(1)}= \sigma^B_{q{\bar q} 
\rightarrow V} \frac{\alpha_s(\mu_R^2)}{\pi} 
\left\{c_3 {\cal D}_1(x) +c_2{\cal D}_0(x) +c_1 \delta(1-x)\right\}
\eeq
with $c_3=4C_F$,
\beq
c_2=-2C_F \ln\left(\frac{\mu_F^2}{Q^2}\right)\, ,
\quad \quad 
c_1^{\mu}=-\frac{3}{2} C_F \ln\left(\frac{\mu_F^2}{Q^2}\right) \, ,
\eeq
which reproduces Eq. (B.3) in Ref. \cite{DY1}, and we identify the non-scale
$\delta(1-x)$ terms as $T_1=2C_F\zeta_2-4C_F$.

At NNLO our master formula, Eq. (\ref{NNLOmaster}), 
reproduces the $D_3(x),D_2(x),D_1(x),D_0(x)$, and $\delta(1-x)$ terms. 
Note that
our results use explicitly the beta function $\beta_0$, thus simplifying the
expression in Eq. (B.8) of \cite{DY1}. We also note that we use the full
NNLO soft-plus-virtual result in \cite{DY1} to derive the two-loop soft 
anomalous dimension, that appears in the $D_0(x)$ term in our master 
formula, for the Drell-Yan process :
\beq
{\rm Re}{\Gamma'}_{S, q\bar q}^{(2)}-{\nu}_q^{(2)}
=C_F C_A \left(\frac{7}{4} \zeta_3
+\frac{11}{3}\zeta_2-\frac{299}{54}\right)+ n_f C_F \left(-\frac{2}{3}\zeta_2
+\frac{25}{27}\right). 
\eeq
Note that this term is universal for all processes
with quark-antiquark annihilation.
We also determine for the Drell-Yan process in the 
$\overline {\rm MS}$ scheme the term $R$ that appears in the 
$\delta(1-x)$ terms in our master formula and which is also universal in 
all $q{\bar q}$ processes:
\beqa
R_{q \bar q}&=&C_F^2\left(-\frac{59}{10}\zeta_2^2
+\frac{29}{8}\zeta_2-\frac{15}{4}\zeta_3
+12 \zeta_4-\frac{1}{64}\right)+C_FC_A\left(-\frac{3}{20}\zeta_2^2
+\frac{37}{9}\zeta_2+\frac{7}{4}\zeta_3-\frac{1535}{192}\right)
\nonumber \\ &&
{}+n_f C_F\left(\frac{\zeta_3}{2}-\frac{7}{9}\zeta_2+\frac{127}{96}\right) \, .
\eeqa

The NNLO soft and virtual corrections for Drell-Yan production
have also been calculated in the DIS scheme in Ref. \cite{DY2}. 
Using our NLO formula in the DIS scheme, Eq. (\ref{NLODIS}),
we find $c'_3=2C_F$,
$c'_2=3C_F/2-2C_F \ln(\mu_F^2/Q^2)$,
${c'}_1^{\mu}=-(3/2) C_F \ln(\mu_F^2/Q^2)$,
and $T'_1=4C_F\zeta_2+C_F/2$, which agress with Eq. (A.3) in \cite{DY2}.

Using our NNLO master formula in the DIS scheme, Eq. (\ref{NNLOd}), we 
are also able to rederive the NNLO result in Eq. (A.8) (plus the 
renormalization scale terms in Eq. (A.7)) of Ref. \cite{DY2},
and thus we identify the $B_q^{(2)}$ term in the DIS scheme master formula: 
\beq
B_q^{(2)}=C_F^2\left(\frac{3}{32}-\frac{3}{4}\zeta_2+\frac{3}{2}\zeta_3\right)
+C_F C_A \left(-\frac{5}{2}\zeta_3+\frac{4937}{864}\right)
-\frac{409}{432}n_f C_F \, .
\eeq
Finally, we also determine for the Drell-Yan process in the 
DIS scheme the term $R'$ that appears in the 
$\delta(1-x)$ term in our master formula:
\beqa
R'_{q \bar q}&=&C_F^2\left(-\frac{43}{20}\zeta_2^2
-\frac{17}{16}\zeta_2+\frac{9}{2}\zeta_3
+3 \zeta_4-\frac{1}{8}\right)+C_FC_A\left(-\frac{77}{40}\zeta_2^2
+\frac{1049}{72}\zeta_2-\frac{49}{12}\zeta_3+\frac{215}{144}\right)
\nonumber \\ &&
{}+n_f C_F\left(\frac{\zeta_3}{3}-\frac{85}{36}\zeta_2-\frac{19}{72}\right)\, .
\eeqa
This is also universal in all $q{\bar q}$ processes in the DIS scheme,
as we will verify below for $W^+ \gamma$ production.

\subsection{Standard Model Higgs production}

Our next application is Higgs production in the Standard Model
in hadron-hadron collisions, for which
the full NNLO corrections, using an effective Lagrangian for the Higgs-gluon 
interaction, have been calculated in
Refs. \cite{Higgs1,Higgs2,Higgs3}. The partonic process we discuss is 
$gg \rightarrow H + X$, where $H$ is the Higgs boson.
At threshold $s=M_H^2$, where $s$ is the center-of-mass energy
squared of the incoming gluon pair and $M_H$ is the Higgs mass. 
Evidently at lowest order there are no final-state massless partons and 
the cross section is given in PIM kinematics.
Here the plus distributions are $D_l(x)$ with $x=M_H^2/s$,
${\rm Re}{\Gamma'}_S^{(1)}=C_A$, and we choose the hard scale
$M^2=M_H^2$.
The soft and virtual corrections for the total cross section 
$\sigma_{gg \rightarrow H}$ are given explicitly 
at NNLO in Refs. \cite{Higgs1,Higgs2,Higgs3,Higgs4,Higgs5}. 
We reproduce and generalize those results by keeping the factorization and 
renormalization scales distinct, using the 
beta function $\beta_0$ explicitly, and keeping the color factors 
$C_A$ explicit in our results.

At NLO, our $\overline{\rm MS}$ scheme master formula gives
\beq
{\hat\sigma}_{gg \rightarrow H}^{(1)}= \sigma^B_{gg \rightarrow H}
\frac{\alpha_s(\mu_R^2)}{\pi} 
\left\{c_3 {\cal D}_1(x) +c_2{\cal D}_0(x)+c_1 \delta(1-x)\right\}
\eeq
with $c_3=4C_A$,
\beq
c_2=-2C_A \ln\left(\frac{\mu_F^2}{M_H^2}\right)\, ,
\quad \quad
c_1^{\mu}=\frac{\beta_0}{2} \ln\left(\frac{\mu_R^2}{\mu_F^2}\right) \, ,
\eeq
which reproduces the results in \cite{Higgs1,Higgs2,Higgs3,Higgs4,Higgs5}. 
We also identify $T_1=11/2+2C_A \zeta_2$.
At NNLO, our master formula reproduces the $D_3(x),D_2(x),D_1(x),D_0(x)$, 
and $\delta(1-x)$ terms. 
We note that we use the full NNLO soft-plus-virtual result in these references
to derive the two-loop soft anomalous dimension,
that appears in the $D_0(x)$ term in our master formula,
for Higgs production:
\beq
{\rm Re}{\Gamma'}_{S,gg}^{(2)}-{\nu}_g^{(2)}=C_A^2 \left(\frac{7}{4} \zeta_3
+\frac{11}{3}\zeta_2-\frac{41}{216}\right)+n_f C_A \left(-\frac{2}{3}\zeta_2
-\frac{5}{108}\right)\, .
\eeq
This anomalous dimension is universal for all processes 
with gluon-gluon fusion.

We also determine for Higgs production in the 
$\overline {\rm MS}$ scheme the term $R$ that appears in the 
$\delta(1-x)$ terms in our master formula and which is also universal in 
all $gg$ processes:
\beq
R_{gg}=\frac{9221}{144}+\frac{67}{2}\zeta_2
-\frac{1089}{20}\zeta_2^2-\frac{165}{4}\zeta_3+108 \, \zeta_4
+n_f \left(-\frac{1189}{144}-\frac{5}{3}\zeta_2
+\frac{5}{6}\zeta_3\right) \, .
\eeq

\subsection{Deep inelastic scattering}

Our methods can also be applied to the coefficient functions in
deep inelastic scattering, $\gamma^* q \rightarrow q$, where the
distributions are $D_l(z)$. We note that here we have a massless parton
(quark) in the final state.
In the ${\overline {\rm MS}}$ scheme, our NLO formula gives: 
$c_3=C_F$, $c_2=-3C_F/4 - C_F \ln(\mu_F^2/Q^2)$, and
$c_1=-(3C_F/4)\ln(\mu_F^2/Q^2)-C_F \zeta_2-9C_F/4$.
The NNLO corrections are then given by our NNLO master formula.
Our NLO and NNLO corrections agree with the results for the 
coefficient functions in Refs. \cite{DIS1,DIS2} (see Appendix B in 
\cite{DIS1} and Appendix A in \cite{DIS2}) and we identify
the two-loop $B'_q$  that appears in the $D_0(x)$ terms in our master
formula:
\beq
{B'}_q^{(2)}=C_F^2\left(\frac{3}{32}-\frac{3}{4}\zeta_2
+\frac{3}{2}\zeta_3\right)
+C_F C_A\left(\zeta_3+\frac{55}{12}\zeta_2+\frac{319}{864}\right)
+n_f C_F \left(-\frac{5}{6}\zeta_2-\frac{59}{432}\right)\, .
\eeq
I have checked that in the DIS scheme the NLO and NNLO corrections
that do not involve the scale vanish, as expected (after all this is the 
definition of the DIS scheme, that the corrections to deep inelastic
scattering in that scheme vanish). This involves checking that
$2 {\Gamma'}_{S, qq}^{(2)}-2\nu_q^{(2)}-{B'}_q^{(2)}+(\beta_0/4) D_q
+B_q^{(2)}=0$ which is a further test of the correctness of the expressions
for the various two-loop quantities that we have derived.

\subsection{$W^+ \gamma$ production}

We now discuss the cross section $s^2 d^2{\hat\sigma}/(dt du)$
for $W^+ \gamma$ production in $p{\bar p}$ collisions for which
the NNLO soft-plus-virtual corrections in the DIS scheme have been
presented in Ref. \cite{MSN}. 
The lowest-order partonic process is $q {\bar q} \rightarrow W^+ \gamma$
and ${\rm Re}{\Gamma'}_S^{(1)}=C_F$.
We define $s_4=s+t+u-m_W^2$, with $s=(p_q+p_{\bar q})^2$,
$t=(p_q-p_{\gamma})^2$, $u=(p_{\bar q}-p_{\gamma})^2$, and $t_1=t-m_W^2$,
$u_1=u-m_W^2$. The plus distributions are $D_l(s_4)$,
and we choose $M^2=s$. 

In the DIS scheme at NLO, our master formula gives
\beq
{\hat\sigma}_{q {\bar q} \rightarrow W^+ \gamma}^{(1)}
= \sigma^B_{q {\bar q} \rightarrow W^+ \gamma} \frac{\alpha_s(\mu_R^2)}{\pi} 
\left\{c'_3 {\cal D}_1(s_4) +c'_2{\cal D}_0(s_4)+c'_1 \delta(s_4)\right\}
\eeq
with $c'_3=2C_F$,
\beq
c'_2=\frac{3}{2}C_F-C_F \ln\left(\frac{t_1u_1}{s^2}\right)
-2C_F\ln\left(\frac{\mu_F^2}{s}\right)\, ,
\eeq
and
\beq
{c'}_1^{\mu}=C_F\left[-\frac{3}{2}+\ln\left(\frac{t_1u_1}{s^2}\right)\right]
\ln\left(\frac{\mu_F^2}{s}\right) \, .
\eeq
Our NLO expansion agrees with Eq. (3.2) in \cite{MSN} and we
identify 
\beq
T'_1=\frac{1}{2}C_F\ln^2\left(\frac{-t_1}{s}\right)
+\frac{1}{2}C_F\ln^2\left(\frac{-u_1}{s}\right)
-\frac{3}{4}C_F \ln\left(\frac{t_1u_1}{s^2}\right)
+4C_F \zeta_2 +\frac{C_F}{2} \, .
\eeq
Using our NNLO master formula in the DIS scheme, we are also able to 
rederive the NNLO soft and virtual corrections
for that process in Eqs. (B1) and (B2) of Ref. \cite{MSN}. 
Since this is the only process previously calculated at NNLO not involving PIM
kinematics, and hence the $\ln(-t_i/M^2)$ terms are explicit,
this re-derivation provides an additional highly non-trivial test of our 
NNLO master formula.

Finally, we note that we can easily derive the NNLO corrections for this
process in the $\overline{\rm MS}$ scheme for which no results have been 
given in the literature. Now, $c_3=4C_F$, $c_2=-2C_F \ln(t_1u_1/s^2)
-2C_F\ln(\mu_F^2/s)$, $c_1^{\mu}={c'}_1^{\mu}$, and
$T_1=C_F\ln^2(-t_1/s)+C_F\ln^2(-u_1/s)+2C_F \zeta_2 -4C_F$.
The NNLO corrections in the $\overline{\rm MS}$ scheme are then given by our
master formula.

\subsection{Charged Higgs production}

We now present new results for processes for which no full NNLO calculations 
have ever been done.

We first consider the production of a charged Higgs boson
in the Minimal Supersymmetric Standard Model, for which there have been
recent NLO calculations \cite{cHiggs1,cHiggs2}.
The lowest-order partonic process is
${\bar b}(p_{\bar b})+ g(p_g) \rightarrow H^+ (p_{H^+})+{\bar t}(p_{\bar t})$.
Again there are no final-state massless partons and in this case
we work in 1PI kinematics. Here ${\hat\sigma}$ can stand, for example, for
$E_{H^+} \; d\sigma/d^3p_{H^+}$.
The relevant hard scale we choose is $M=m_{H^+}$ and we 
define the Mandelstam invariants $s=(p_{\bar b}+p_g)^2$, 
$t=(p_{H^+}-p_{\bar b})^2$, and $u=(p_{H^+}-p_g)^2$.
The threshold variable is $s_2=s+t+u-m_{H^+}^2-m_{\bar t}^2-m_{\bar b}^2$
and the plus distributions are $D_l(s_2)$.
Also, $t_g=t_1=t-m_t^2$ and $t_{\bar b}=u_1=u-m_t^2$.
The one-loop soft anomalous dimension is
${\rm Re}{\Gamma'}_S^{(1)}=C_F[\ln(-u_1/s)+(1/2) \ln(s/m_t^2)]
+(C_A/2)[\ln(t_1/u_1)+1]$.

The NLO soft-plus-virtual corrections in the $\overline{\rm MS}$ scheme are
\beq
{\hat\sigma}_{{\bar b} g \rightarrow H^+ {\bar t}}^{(1)}= 
\sigma^B_{{\bar b} g \rightarrow H^+ {\bar t}} \frac{\alpha_s(\mu_R^2)}{\pi}
\left\{c_3 {\cal D}_1(s_2)+c_2{\cal D}_0(s_2) +c_1 \delta(s_2)\right\}
\eeq
with $c_3=2(C_F+C_A)$,
\beq
c_2=C_F\left[\ln\left(\frac{m_H^4}{sm_t^2}\right)-1
-\ln\left(\frac{\mu_F^2}{s}\right)\right]
+C_A\left[\ln\left(\frac{m_H^4}{t_1 u_1}\right)
-\ln\left(\frac{\mu_F^2}{s}\right)\right] \, ,
\eeq
and
\beq
c_1^{\mu}=\left[C_F \ln\left(\frac{-u_1}{m_H^2}\right)
+C_A \ln\left(\frac{-t_1}{m_H^2}\right)
-\frac{3C_F}{4}-\frac{\beta_0}{4}\right]\ln\left(\frac{\mu_F^2}{s}\right)
+\frac{\beta_0}{4} \ln\left(\frac{\mu_R^2}{s}\right) \, .
\eeq
The NNLO soft and virtual corrections are then explicitly given by our 
master formula.

\subsection{$W,Z$ plus jet production}

$W,Z$ plus jet production in hadron colliders has been studied
at NLO in Refs. \cite{Wjnlo1,Wjnlo2}, and a resummed cross section and
NNLO-NNLL corrections have been presented in Ref. \cite{KV}.
Here we follow the notation of Ref. \cite{KV} and 
discuss the 1PI cross section $E_Q \, d\sigma/d^3 Q$ with $Q$ the momentum
of the electroweak boson.
At lowest order there are two partonic processes,
$q(p_a)+g(p_b)\rightarrow q(p_c)+V(Q)$ and 
$q(p_a)+{\bar q}(p_b)\rightarrow g(p_c)+V(Q)$, where $V$ stands
for $W$ or $Z$. We note that we have final-state massless partons in 
both processes. We define the kinematic invariants
$s=(p_a+p_b)^2$, $t=(p_a-Q)^2$, and $u=(p_b-Q)^2$.
We choose $M^2=Q^2$, and the plus distributions are $D_l(s_2)$
with $s_2=s+t+u-Q^2$.
We discuss the $\overline{\rm MS}$ corrections for the two 
partonic processes in turn. 

\subsubsection{$q {\bar q} \rightarrow g V$}

Here $t_q=u$, $t_{\bar q}=t$, 
and ${\rm Re}{\Gamma'}_S^{(1)}=C_F+(C_A/2) \ln(tu/s^2)+C_A/2$ \cite{KV}.
The NLO soft and virtual corrections are 
\beq
{\hat\sigma}_{q{\bar q} \rightarrow gV}^{(1)}= 
\sigma^B_{q{\bar q} \rightarrow gV} \frac{\alpha_s(\mu_R^2)}{\pi}
\left\{c_3 {\cal D}_1(s_2)+c_2 {\cal D}_0(s_2)+c_1 \delta(s_2) \right\}
\eeq
with $c_3=4 C_F-C_A$,
\beq
c_2=-\frac{\beta_0}{4}-2C_F \ln\left(\frac{\mu_F^2}{Q^2}\right)
-(2C_F-C_A)\ln\left(\frac{tu}{sQ^2}\right)\, ,
\eeq
and
\beq
c_1^{\mu}=C_F\left[-\frac{3}{2}+\ln\left(\frac{tu}{Q^4}\right)\right]
\ln\left(\frac{\mu_F^2}{s}\right) 
+\frac{\beta_0}{4} \ln\left(\frac{\mu_R^2}{s}\right)\, .
\eeq
These are in agreement with the NLO result in \cite{Wjnlo2}.

\subsubsection{$q g \rightarrow q V$}

Here $t_q=u$, $t_g=t$, and ${\rm Re}{\Gamma'}_S^{(1)}=C_F \ln(-u/s)
+C_F+(C_A/2)\ln(t/u)+C_A/2$ \cite{KV}.
The NLO soft and virtual corrections are
\beq
{\hat\sigma}_{qg \rightarrow qV}^{(1)}= \sigma^B_{qg \rightarrow qV}
\frac{\alpha_s(\mu_R^2)}{\pi} \left\{c_3 {\cal D}_1(s_2)+c_2 {\cal D}_0(s_2) 
+c_1 \delta(s_2) \right\}
\eeq
with $c_3=C_F+2C_A$, 
\beq
c_2=-\frac{3}{4}C_F-(C_F+C_A)\ln\left(\frac{\mu_F^2}{Q^2}\right)
-C_A\ln\left(\frac{tu}{sQ^2}\right)\, ,
\eeq
and
\beq
c_1^{\mu}=\left[-\frac{\beta_0}{4}-\frac{3}{4}C_F
+C_F\ln\left(\frac{-u}{Q^2}\right)+C_A\ln\left(\frac{-t}{Q^2}\right)\right]
\ln\left(\frac{\mu_F^2}{s}\right) 
+\frac{\beta_0}{4} \ln\left(\frac{\mu_R^2}{s}\right)\, .
\eeq
They are in agreement with the NLO result in \cite{Wjnlo2}.

For both partonic subprocesses the NNLO corrections are derived from
our master formula and are in agreement with the NNLO-NNLL results
in Ref. \cite{KV} (apart from a term $\sigma^B [\alpha_s^2(\mu_R^2)/\pi^2]
(\beta_0/4) c_3 \ln(Q^2/s) {\cal D}_1$ 
that was missing in that reference).

\subsection{Direct photon production}

Direct photon production is often recognised as a process that can
aid determinations of the gluon distribution.
The NLO cross section for direct photon production has been
given in Refs. \cite{ADBFS,GV}.  
At lowest order, the parton-parton scattering  subprocesses are 
$ q(p_a)+g(p_b) \rightarrow  \gamma(p_{\gamma}) + q(p_J)$
and $q(p_a)+{\bar q}(p_b) \rightarrow \gamma(p_{\gamma}) + g(p_J)$,
so there are final-state massless partons in both subprocesses. 
We define the Mandelstam invariants 
$s=(p_a+p_b)^2$, $t=(p_a-p_{\gamma})^2$, and 
$u=(p_b-p_{\gamma})^2$,
which satisfy $s_4 \equiv s+t+u=0$ at threshold.
Here we choose $M^2=p_T^2=tu/s$, and we work in 1PI kinematics 
in the $\overline {\rm MS}$ scheme with the 
cross section $E_{\gamma} \; d^3\sigma/d^3 p_{\gamma}$.
The threshold logarithms $D_l(s_4)$ have been resummed and 
NNLO-NNLL corrections have been presented in Ref. \cite{KO1}.

\subsubsection{$q {\bar q} \rightarrow \gamma g$}

We start with the process $q {\bar q} \rightarrow \gamma g$
for which $t_q=u$, $t_{\bar q}=t$, and 
${\rm Re}{\Gamma'}_S^{(1)}=C_F+(C_A/2) \ln(tu/s^2)+C_A/2$. 
The NLO soft plus virtual corrections are
\beq
{\hat\sigma}_{q {\bar q} \rightarrow \gamma g}^{(1)}= 
\sigma^B_{q {\bar q} \rightarrow \gamma g} \frac{\alpha_s(\mu_R^2)}{\pi}
\left\{c_3 {\cal D}_1(s_4) +c_2 {\cal D}_0(s_4)+c_1 \delta(s_4) \right\}
\eeq
with
$c_3=4 C_F-C_A$,
\beq
c_2=-\frac{\beta_0}{4}-2C_F \ln\left(\frac{\mu_F^2}{p_T^2}\right)\, ,
\eeq
and $c_1=c_1^{\mu}+T_1$ with
\beq
c_1^{\mu}=C_F\left[-\frac{3}{2}-\ln\left(\frac{p_T^2}{s}\right)\right]
\ln\left(\frac{\mu_F^2}{s}\right) 
+\frac{\beta_0}{4} \ln\left(\frac{\mu_R^2}{s}\right).
\eeq
The term $T_1$ is given in Eq. (3.11) of Ref. \cite{KO1} 
(where it is called ${c'}_1^{q{\bar q}}$).
The NNLO soft plus virtual corrections are then explicitly given 
by our master formula.
We note that the terms $D_3(s_4)$, $D_2(s_4)$, and $D_1(s_4)$
were already derived from a resummation study in Section III of 
Ref. \cite{KO1} and are in agreement with our formula.
   
\subsubsection{$q g \rightarrow \gamma q$}

We continue with the process $q g \rightarrow \gamma q$
for which $t_q=u$, $t_g=t$, and
${\rm Re}{\Gamma'}_S^{(1)}=C_F \ln(-u/s)+C_F+(C_A/2)\ln(t/u)+C_A/2$.
The NLO soft plus virtual corrections are
\beq
{\hat\sigma}_{q g \rightarrow \gamma q}^{(1)}= 
\sigma^B_{q g \rightarrow \gamma q} \frac{\alpha_s(\mu_R^2)}{\pi}
\left\{c_3 {\cal D}_1(s_4)+c_2 {\cal D}_0(s_4)+c_1 \delta(s_4) \right\}
\eeq
with
$c_3=C_F+2C_A$,
\beq
c_2=-\frac{3}{4}C_F-(C_F+C_A)\ln\left(\frac{\mu_F^2}{p_T^2}\right)\, ,
\eeq
and $c_1=c_1^{\mu}+T_1$ with
\beq
c_1^{\mu}=\left[-\frac{\beta_0}{4}-\frac{3}{4}C_F
-C_F\ln\left(\frac{-t}{s}\right)-C_A\ln\left(\frac{-u}{s}\right)\right]
\ln\left(\frac{\mu_F^2}{s}\right) 
+\frac{\beta_0}{4} \ln\left(\frac{\mu_R^2}{s}\right)\, .
\eeq
The term $T_1$ is given in Eq. (3.8) of Ref. \cite{KO1} 
(where it is called ${c'}_1^{qg}$).
The NNLO soft plus virtual corrections are then explicitly given 
by our master formula. Again,
we note that the terms $D_3(s_4)$, $D_2(s_4)$, and $D_1(s_4)$
were already derived from a resummation study in Section III of
Ref. \cite{KO1} and are in agreement with our formula.

\subsection{DIS heavy quark production}

The NNLO corrections to heavy quark production in deep inelastic 
scattering may be needed, together with various resummations \cite{ELSM,KNOY}, 
in explaining the discrepancy between NLO theory \cite{LRS} and experiment
for bottom quark production.
Here the lowest-order partonic process is
$\gamma^* g \rightarrow Q {\bar Q}$, so there are no final-state 
massless partons, and we are working in 1PI kinematics 
in the $\overline{\rm MS}$ scheme with the cross section
$d^2{\sigma}/(dt_1 du_1)$.
The soft anomalous dimension is
${\rm Re}{\Gamma'}_S^{(1)}=(C_A/2-C_F) ({\rm Re} L_{\beta}+1)
+(C_A/2)\ln(t_1 u_1/(m^2s))$, where $s=(p_{\gamma^*}+p_g)^2$,
$t_1=(p_g-p_Q)^2-m^2$, and $u_1=(p_{\gamma^*}-p_Q)^2-m^2$,
with $m$ the heavy quark mass.
The singular distributions are $D_l(s_2)$ with $s_2=s+t_1+u_1$,
and we use $M=m$.

The NLO corrections are
\beq
{\hat\sigma}_{\gamma^* g \rightarrow Q {\bar Q}}^{(1)}= 
\sigma^B_{\gamma^* g \rightarrow Q {\bar Q}} \frac{\alpha_s(\mu_R^2)}{\pi}
\left\{c_3 {\cal D}_1(s_2)+c_2 {\cal D}_0(s_2) +c_1 \delta(s_2) \right\}
\eeq
with $c_3=2C_A$,
\beq
c_2=-2C_F({\rm Re} L_{\beta}+1)+C_A\left[{\rm Re} L_{\beta}
+\ln\left(\frac{t_1}{u_1}\right)-\ln\left(\frac{\mu_F^2}{m^2}\right) 
\right]\, ,
\eeq
and
\beq
c_1^{\mu}=\left[-\frac{\beta_0}{4}
+C_A\ln\left(\frac{-u_1}{m^2}\right)\right]
\ln\left(\frac{\mu_F^2}{s}\right) 
+\frac{\beta_0}{4} \ln\left(\frac{\mu_R^2}{s}\right) \, .
\eeq
The NNLO soft and virtual corrections are then given by our master 
formula. We note that the $D_3(s_2)$ and $D_2(s_2)$ terms were derived in
Ref. \cite{ELSM} (with $\mu_F=\mu_R$) and are in agreement with our results. 

\subsection{FCNC single-top production}

The last process with simple color flow that we consider is single-top
production mediated by flavor-changing neutral currents~(FCNC).
The QCD corrections to the 1PI cross section $d\sigma/(dt \, du)$
in the $\overline{\rm MS}$ scheme for FCNC single-top-quark production 
in $ep$ collisions at the HERA collider 
were studied at NLO in the eikonal approximation in Ref. \cite{BK}.
Here the partonic process is $eu \rightarrow et$, so 
there are no final-state massless partons, and
${\rm Re}{\Gamma'}_S^{(1)}=C_F \ln[(m_t^2-t)/(\sqrt{s}m_t)]$, with
$s=(p_e+p_u)^2$, $t=(p_t-p_u)^2$, and $u=(p_t-p_e)^2$.
The singular distributions are $D_l(s_2)$ with $s_2=s+t+u-m_t^2-2m_e^2$,
and we choose $M=m_t$.

The NLO corrections are
\beq
{\hat\sigma}_{eu \rightarrow et}^{(1)}= \sigma^B_{eu \rightarrow et}
\frac{\alpha_s(\mu_R^2)}{\pi} \left\{c_3 {\cal D}_1(s_2)+c_2 {\cal D}_0(s_2) 
+c_1 \delta(s_2) \right\}
\eeq
with $c_3=2C_F$,
\beq
c_2=C_F\left[-1-2\ln\left(\frac{-u+m_e^2}{m_t^2}\right)
+2\ln\left(\frac{m_t^2-t}{m_t^2}\right)
-\ln\left(\frac{\mu_F^2}{m_t^2}\right)\right]\, ,
\eeq
and
\beq
c_1^{\mu}=\left[-\frac{3}{4} 
+\ln\left(\frac{-u+m_e^2}{m_t^2}\right)\right]
C_F\ln\left(\frac{\mu_F^2}{s}\right) \, .
\eeq
These agree with the results in \cite{BK}.
The NNLO corrections are then given by our master formula.

\mysection{NNLO master formula and applications 
- complex color flow}

\subsection{NLO and NNLO master formulas for complex color flow}

When the lowest-order cross section involves already a complex color flow
that is expressed in terms of non-trivial color matrices, then the 
master formulas
given in Section 2 have to be extended. 
Now not all the NLO corrections are proportional to the
Born term; only the leading logarithms and terms involving the scale are.
Therefore at NLO our master formula for soft and virtual corrections
in the $\overline {\rm MS}$ scheme is extended for the case of complex color 
flow as
\beq
{\hat{\sigma}}^{(1)} = \sigma^B \frac{\alpha_s(\mu_R^2)}{\pi}
\left\{c_3\, {\cal D}_1(x_{th}) + c_2\,  {\cal D}_0(x_{th}) 
+c_1\,  \delta(x_{th})\right\}+\frac{\alpha_s^{d_{\alpha_s}+1}(\mu_R^2)}{\pi} 
\left[A^c \, {\cal D}_0(x_{th})+T_1^c \, \delta(x_{th})\right] \, ,
\label{NLOcx}
\eeq
where $c_3$ is defined as before in Eq. (\ref{c3}), 
and $c_2$ is defined by 
\beq
c_2=- \sum_i \left[C_{f_i}
+2 C_{f_i} \, \delta_K \, \ln\left(\frac{-t_i}{M^2}\right)+
C_{f_i} \ln\left(\frac{\mu_F^2}{s}\right)\right]
-\sum_j \left[{B'}_j^{(1)}+C_{f_j}
+C_{f_j} \, \delta_K \, \ln\left(\frac{M^2}{s}\right)\right] \, ,
\label{c2n}
\eeq
i.e. is the same as in the simple color flow case except without the
term $2{\rm Re}{\Gamma'}_S^{(1)}$. 
The function $A^c$
is process-dependent and depends on the color structure of the hard-scattering,
for which specific examples are given in the next 
subsections. It is defined by
\beq
A^c={\rm Tr} \left(H^{(0)} {\Gamma'}_S^{(1)\,\dagger} S^{(0)}
+H^{(0)} S^{(0)} {\Gamma'}_S^{(1)}\right) \, .
\label{Ac}
\eeq
Note that we use the expansions for the hard and soft matrices:
$H=\alpha_s^{d_{\alpha_s}}H^{(0)}+(\alpha_s^{d_{\alpha_s}+1}/\pi)H^{(1)}
+(\alpha_s^{d_{\alpha_s}+2}/\pi^2)H^{(2)}+\cdots$ and
$S=S^{(0)}+(\alpha_s/\pi)S^{(1)}+(\alpha_s/\pi)^2 S^{(2)}+\cdots$.
The Born term is then given by $\sigma^B={\rm Tr}[H^{(0)}S^{(0)}]$.

With respect to the $\delta(x_{th})$ terms, we split them into a term
$c_1=c_1^{\mu}+T_1$, with $c_1^{\mu}$ defined in Eq. (\ref{c1mu})
as before, that is proportional 
to the Born cross section, and a term $T_1^c$ that is not.
$T_1^c$ is also process-dependent, is formally defined by 
\beqa
T_1^c&=&{\rm Tr} \left(H^{(1)}S^{(0)}+H^{(0)}S^{(1)}\right)
+A^c \, \delta_K \, \ln\left(\frac{M^2}{s}\right)
+\sigma^B \frac{\alpha_s(\mu_R^2)}{\pi} \left[-T_1 +\frac{c_3}{2}
\delta_K \ln^2\left(\frac{M^2}{s}\right) \right.
\nonumber \\ && \left.
{}+T_2 \, \delta_K \ln\left(\frac{M^2}{s}\right) \right]\, ,
\eeqa
and can also be derived by matching to a full NLO calculation.

In the DIS scheme the corresponding terms $c_3'$, $c_2'$, and $c_1'$ are 
given as in Section 2.2.2.

At NNLO the master formula for soft and virtual corrections 
in the $\overline {\rm MS}$ scheme is extended for the case of complex 
color flow as
\beqa
{\hat{\sigma}}^{(2)}&=&{\hat{\sigma}}^{(2)}_{\rm simple}
+\frac{\alpha_s^{d_{\alpha_s}+2}(\mu_R^2)}{\pi^2} 
\left\{\frac{3}{2} c_3 \, A^c\, {\cal D}_2(x_{th})
+\left[\left(2c_2-\frac{\beta_0}{2}\right)A^c+c_3 T_1^c
+F\right] {\cal D}_1(x_{th})
\right.
\nonumber \\ && \hspace{-20mm}
{}+\left[\left(c_1-\zeta_2 c_3+\frac{\beta_0}{4}
\ln\left(\frac{\mu_R^2}{s}\right)\right)A^c+\left(c_2-\frac{\beta_0}{2}\right)
T_1^c+F \, \delta_K \, \ln\left(\frac{M^2}{s}\right)+G\right] 
{\cal D}_0(x_{th})
\nonumber \\ && \hspace{-20mm} 
{}+\left[(\zeta_3 c_3-\zeta_2 c_2)A^c+\left(c_1+\frac{\beta_0}{4}
\ln\left(\frac{\mu_R^2}{s}\right)\right)T_1^c 
+\frac{1}{2}\left(\delta_K \, \ln^2\left(\frac{M^2}{s}\right)-\zeta_2\right)F
\right.
\nonumber \\ && \hspace{-20mm} \quad \left. \left.
{}-\frac{\beta_0}{2} \, \delta_K \, T_1^c \ln\left(\frac{M^2}{s}\right)
+\frac{\beta_0}{4} \, \delta_K \, A^c \ln^2\left(\frac{M^2}{s}\right)
+G \, \delta_K \, \ln\left(\frac{M^2}{s}\right)+R^c \right]
\delta(x_{th})\right\} \, .
\nonumber \\
\label{NNLOc}
\eeqa
Here ${\hat{\sigma}}^{(2)}_{\rm simple}$ denotes the expression
in Eq. (\ref{NNLOm}) after using the new $c_2$ of Eq. (\ref{c2n})
everywhere in that expression and deleting all ${\Gamma'}_S$, and $R$, 
from that expression. Also, we have used 
\beq
F={\rm Tr} \left[H^{(0)} \left({\Gamma'}_S^{(1)\,\dagger}\right)^2 S^{(0)}
+H^{(0)} S^{(0)} \left({\Gamma'}_S^{(1)}\right)^2
+2 H^{(0)} {\Gamma'}_S^{(1)\,\dagger} S^{(0)} {\Gamma'}_S^{(1)} \right] \, ,
\label{Fterm}
\eeq
\beqa
G&=&{\rm Tr} \left[H^{(1)} {\Gamma'}_S^{(1)\,\dagger} S^{(0)}
+H^{(1)} S^{(0)} {\Gamma'}_S^{(1)} + H^{(0)} {\Gamma'}_S^{(1)\,\dagger} S^{(1)}
+H^{(0)} S^{(1)} {\Gamma'}_S^{(1)} \right.
\nonumber \\ && \quad \quad \left.
{}+H^{(0)} {\Gamma'}_S^{(2)\,\dagger} S^{(0)}
+H^{(0)} S^{(0)} {\Gamma'}_S^{(2)} \right] \, ,
\eeqa
and
\beq
R^c={\rm Tr} \left(H^{(2)}S^{(0)}+H^{(0)} S^{(2)}+H^{(1)} S^{(1)}\right)
-\frac{1}{2} T_1^2 -T_1 T_1^c \, .
\eeq

I have checked explicitly that the renormalization and 
factorization scale dependence cancels in the physical cross section at NNLO.

In the DIS scheme the NNLO corrections are given by Eq. (\ref{NNLOd}),
after replacing the term $\sigma^B (\alpha_s^2(\mu_R^2)/\pi^2)
{\hat{\sigma'}}^{(2)}|_{c_i'}$ by the cross section in Eq. (\ref{NNLOc}), 
having also replaced in Eq. (\ref{NNLOc}) all the $c_i$,
$T_i$, $G$, and $R^c$, by their DIS counterparts $c_i'$, $T_i'$, 
$G'$, and ${R^c}'$.

We now apply the NLO and NNLO master formulas to a variety of processes
with complex color flow.

\subsection{Heavy quark hadroproduction}

The production cross sections of top, bottom, and charm quarks in 
hadron colliders can be considerably enhanced near threshold.
The NLL resummation of threshold corrections was derived in Ref. \cite{KS} 
and the NNLO-NNLL corrections for heavy quark total cross sections and 
differential distributions were calculated in Refs. \cite{NK,KLMV} in both 
1PI and PIM kinematics. There are two partonic channels involved at lowest
order, $q{\bar q} \rightarrow Q {\bar Q}$ and $gg \rightarrow Q {\bar Q}$.
We choose the hard scale $M=m$, with $m$ the heavy quark mass.
The 1PI cross section is $s^2\, d^2{\hat\sigma}/(dt_1 \, du_1)$,
with $t_1=t-m^2$, $u_1=u-m^2$, and the singular distributions
are ${\cal D}_l(s_4)$ with $s_4=s+t_1+u_1$.
The PIM cross section is $s \, d^2{\hat\sigma}/(dM_{Q\bar Q}^2 
d\cos\theta)$, with $M_{Q\bar Q}$ the heavy quark pair mass and $\theta$ 
the scattering
angle in the partonic center-of-mass frame, and the singular distributions
are ${\cal D}_l(z)$ with $z=M_{Q\bar Q}^2/s$.
The one-loop soft anomalous dimension matrices $\Gamma_S^{(1)}$ 
were calculated for both partonic processes in Ref. \cite{KS}. 
Explicit results for the matrices $H^{(0)}$ and $S^{(0)}$ can be found in
Refs. \cite{NK,KLMV} for both partonic channels.

\subsubsection{$q {\bar q} \rightarrow Q {\bar Q}$ channel}

We begin with the quark-antiquark annihilation channel.
We note that $\Gamma_S'$, $H$, and $S$ are $2 \times 2$ matrices.
However, the triviality of the hard matrix $H^{(0)}$ (only one non-zero
element) leads to almost simple-color-flow-like expressions.

The NLO soft plus virtual corrections in 1PI kinematics in the 
$\overline{\rm MS}$ scheme are given by Eq. (\ref{NLOcx}) with $c_3=4C_F$,
\beq
c_2=-2C_F-2C_F\ln\left(\frac{t_1u_1}{m^4}\right)
-2C_F\ln\left(\frac{\mu_F^2}{s}\right)\, ,
\eeq
\beq
c_1^{\mu}=C_F \left[\ln\left(\frac{t_1u_1}{m^4}\right)-\frac{3}{2}\right]
\ln\left(\frac{\mu_F^2}{s}\right) 
+\frac{\beta_0}{2}\ln\left(\frac{\mu_R^2}{s}\right)\, ,
\eeq
and $A^c=(\sigma^B/ \alpha_s^2) \, 2\, {\rm Re}{\Gamma'}_{S,22}^{(1)}$,
and are in agreement with the NLO results in \cite{NLOqq}.
Here the real part of the one-loop $22$ element of the soft anomalous 
dimension matrix is ${\rm Re}{\Gamma'}_{S,22}^{(1)}=C_F[4\ln(u_1/t_1)
-{\rm Re} L_{\beta}]+(C_A/2)[-3\ln(u_1/t_1)
-\ln(m^2s/(t_1u_1))+{\rm Re} L_{\beta}]$ with
$L_{\beta}=(1-2m^2/s)/\beta\cdot[\ln((1-\beta)/(1+\beta))+\pi i]$
and $\beta=\sqrt{1-4m^2/s}$.
Explicit expressions for $T_1$, $T_1^c$ can be extracted from 
Ref. \cite{NLOqq}.

The corresponding NLO $\overline{\rm MS}$ corrections in PIM kinematics  
are given by similar expressions after striking out the $\ln(t_1u_1/m^4)$ 
terms from the 1PI $c_2$ and $c_1^{\mu}$, as per our NLO master formula, 
and using the relevant PIM Born term and $T_1$, $T_1^c$ \cite{KLMV}.

The NNLO soft and virtual corrections are then given by our master 
formula for complex color flows for either kinematics and are in agreement
with the NNLO-NNLL results in \cite{NK,KLMV}. We note that the term $F$ in 
Eq. (\ref{Fterm}) has a relatively simple form, 
$F=(\sigma^B/ \alpha_s^2)
[4 ({\rm Re}{\Gamma'}_{S,22}^{(1)})^2+4 {\Gamma'}_{S,12}^{(1)} 
{\Gamma'}_{S,21}^{(1)}]$ with ${\Gamma'}_{S,12}^{(1)}=(C_F/C_A)
\ln(u_1/t_1)$ and ${\Gamma'}_{S,21}^{(1)}=2\ln(u_1/t_1)$.

In the DIS scheme, we have in 1PI kinematics
$c'_3=2C_F$,
\beq
c'_2=-\frac{C_F}{2}-C_F\ln\left(\frac{t_1u_1}{m^4}\right)
-2C_F\ln\left(\frac{\mu_F^2}{s}\right)\, ,
\eeq
\beq
c'_1=c_1-\frac{1}{2}C_F \left[\ln^2\left(\frac{-t_1}{m^2}\right)
+\ln^2\left(\frac{-u_1}{m^2}\right)\right]-\frac{3}{4} C_F
\ln\left(\frac{t_1 u_1}{m^4}\right)+2C_F \zeta_2+\frac{9}{2}C_F.
\eeq 
The corresponding corrections in PIM kinematics  are again given by 
similar expressions after striking out the  $\ln^2(-t_1/m^2)$,
$\ln^2(-u_1/m^2)$, and $\ln(t_1u_1/m^4)$ terms from 
the 1PI $c'_2$ and $c'_1$, 
and using the relevant PIM Born term and ${T_1^c}'$ \cite{KLMV}.
The NNLO soft and virtual corrections in the DIS scheme are then given by 
our master formula for either kinematics.

\subsubsection{$gg \rightarrow Q {\bar Q}$ channel}

We continue with the $gg \rightarrow Q {\bar Q}$ channel whose
color structure is considerably more complex.
Here $\Gamma_S'$, $H$, and $S$ are $3 \times 3$ matrices.
The NLO soft plus virtual corrections are given in 1PI kinematics
in the $\overline{\rm MS}$ scheme
by Eq. (\ref{NLOcx}) with $c_3=4C_A$,
\beq
c_2=-2C_A-2C_A\ln\left(\frac{t_1u_1}{m^4}\right)
-2C_A\ln\left(\frac{\mu_F^2}{s}\right)\, ,
\eeq
\beq
c_1^{\mu}=\left[C_A \ln\left(\frac{t_1u_1}{m^4}\right)-\frac{\beta_0}{2}\right]
\ln\left(\frac{\mu_F^2}{s}\right) 
+\frac{\beta_0}{2}\ln\left(\frac{\mu_R^2}{s}\right)\, ,
\eeq
and
\beqa
A^c&=& \pi K_{gg} B_{QED} (N_c^2-1)\left\{N_c\frac{t_1^2+u_1^2}{s^2}
\left[\left(-C_F+\frac{C_A}{2}\right)({\rm Re}L_{\beta}+1)
+\frac{N_c}{2}+\frac{N_c}{2} \ln\left(\frac{t_1u_1}{m^2 s}\right)\right]
\right.
\nonumber \\ && \left.
{}+\frac{1}{N_c}(C_F-C_A)({\rm Re}L_{\beta}+1)
-\ln\left(\frac{t_1u_1}{m^2 s}\right)
+\frac{N_c^2}{2} \frac{t_1^2-u_1^2}{s^2} \ln\left(\frac{u_1}{t_1}\right)
\right\}\, ,
\eeqa
and are in agreement with the NLO results in \cite{NLOgg}.
Explicit expressions for $T_1$, $T_1^c$ can be extracted from 
Ref. \cite{NLOgg}.
As for the $q{\bar q}$ channel, the corresponding NLO corrections in 
PIM kinematics are obtained after striking out the $\ln(t_1u_1/m^4)$ 
terms from the 1PI $c_2$ and $c_1^{\mu}$ (note that $A^c$ is not affected 
by the kinematics choice) and using the relevant PIM Born term 
and $T_1$, $T_1^c$ \cite{KLMV}.

The term $F$ can be explicitly calculated using Eq. (\ref{Fterm}).
The NNLO soft and virtual corrections for either kinematics 
are then given by our master formula for complex color flow
and are in agreement with the NNLO-NNLL results in \cite{NK,KLMV}.

\subsection{Jet production}

Threshold resummation for jet production has been studied
in Refs. \cite{KOS,KO2}. There are many partonic subprocesses for which 
explicit results for the NNLO corrections at NLL accuracy were given
in \cite{KO2}. Here we are able to extend those results by employing
our master formula. We note that the color structure for these
processes is quite complex, and the one-loop soft anomalous dimension matrices
\cite{KOS,NKrev} along with the lowest-order hard and soft matrices 
can be found in Ref. \cite{KO2}. 
Recently, the complete two-loop virtual corrections
have been calculated \cite{WGQCD,QCDSM,NNLOj1,NNLOj2}. 

Here we discuss the single-jet inclusive cross section
$E_J \, d^3{\hat\sigma}/d^3 p_J$ in the $\overline{\rm MS}$ scheme.
The NLO soft and virtual corrections with $M^2=p_T^2=tu/s$ 
and $s_4=s+t+u$ can be written for each subprocess as 
\beq
{\hat\sigma}^{(1)}=\sigma^B \frac{\alpha_s(\mu_R^2)}{\pi}
\left\{c_3 {\cal D}_1(s_4) + c_2  {\cal D}_0(s_4) 
+c_1 \delta(s_4)\right\}
+\frac{\alpha_s^{d_{\alpha_s}+1}(\mu_R^2)}{\pi}\left[ A^c \, {\cal D}_0(s_4)
+T_1^c \, \delta(s_4)\right] \, .
\label{jetqq}
\eeq
The expressions are symmetric in $t$ and $u$ except for the 
$qg \rightarrow qg$ channel. Full NLO expressions have been given in 
\cite{NLOj}.
We discuss next the individual partonic processes in jet production.

\subsubsection{$q {\bar q} \rightarrow q {\bar q}$ and $qq \rightarrow qq$}

There are several processes involving distinct or identical quarks
and antiquarks: $q_j {\bar q}_j \rightarrow q_j {\bar q}_j$,
$q_j {\bar q}_j \rightarrow q_k {\bar q}_k$,
$q_j {\bar q}_k \rightarrow q_j {\bar q}_k$,
$q_j q_j \rightarrow q_j q_j$,
$q_j q_k \rightarrow q_j q_k$, and the corresponding ones with antiquarks. 

The NLO soft and virtual corrections are given by Eq. (\ref{jetqq})
with $c_3=2C_F$, 
\beq
c_2=-2C_F \ln\left(\frac{\mu_F^2}{s}\right)-\frac{11}{2}C_F \, ,
\eeq
\beq
c_1^{\mu}=-C_F\left[\ln\left(\frac{p_T^2}{s}\right)+\frac{3}{2}\right]
\ln\left(\frac{\mu_F^2}{s}\right)
+\frac{\beta_0}{2}\ln\left(\frac{\mu_R^2}{s}\right) \, .
\eeq
The expressions for $\sigma^B$ and $A^c$ depend on the specific 
process. The expressions for $\sigma^B$ can be found in Appendix A 
(for $q {\bar q} \rightarrow q {\bar q}$ processes) 
and Appendix B (for $qq \rightarrow qq$ processes) 
of Ref. \cite{KO2}.
The expressions for $A^c$ can be easily derived from Eq. (\ref{Ac})
or by comparing Eq. (\ref{jetqq}) with the expressions in Appendix A or
Appendix B of \cite{KO2} for the NLO corrections 
for the various subprocesses.

\subsubsection{$q {\bar q} \rightarrow  gg$ and $gg \rightarrow q {\bar q}$}

The NLO soft and virtual corrections are given by Eq. (\ref{jetqq})
with $c_3=4C_F-2C_A$, 
\beq
c_2=-2C_F\ln\left(\frac{\mu_F^2}{p_T^2}\right)
-\frac{\beta_0}{2}-2C_F-2C_A-2C_A\ln\left(\frac{p_T^2}{s}\right)\, ,
\eeq
\beq
c_1^{\mu}=-C_F\left[\ln\left(\frac{p_T^2}{s}\right)+\frac{3}{2}\right]
\ln\left(\frac{\mu_F^2}{s}\right)
+\frac{\beta_0}{2}\ln\left(\frac{\mu_R^2}{s}\right)\, ,
\eeq
for the process $q {\bar q} \rightarrow  gg$, and
$c_3=4C_A-2C_F$,
\beq
c_2=-2C_A\ln\left(\frac{\mu_F^2}{p_T^2}\right)
-\frac{7}{2}C_F-2C_A-2C_F\ln\left(\frac{p_T^2}{s}\right)\, ,
\eeq
\beq
c_1^{\mu}=-C_A \ln\left(\frac{p_T^2}{s}\right)
\ln\left(\frac{\mu_F^2}{s}\right)
+\frac{\beta_0}{2}\ln\left(\frac{\mu_R^2}{\mu_F^2}\right) \, ,
\eeq
for the process $gg \rightarrow q {\bar q}$.
The expressions for $\sigma^B$  depend on the specific 
process and can be found in Appendix C of Ref. \cite{KO2}.
The expressions for $A^c$ can be easily derived from Eq. (\ref{Ac})
or by comparing
Eq. (\ref{jetqq}) with the expressions in Appendix C of \cite{KO2}
for the NLO corrections for these subprocesses.  

\subsubsection{$q g \rightarrow  q g$}

The NLO soft and virtual corrections, using $t_q=u$ and $t_g=t$,
are given by Eq. (\ref{jetqq})
with $c_3=C_F+C_A$,
\beq
c_2=-(C_F+C_A)\ln\left(\frac{\mu_F^2}{p_T^2}\right)
-\frac{11}{4}C_F-2C_A-\frac{\beta_0}{4}
-2C_F\ln\left(\frac{-u}{s}\right)
-2C_A\ln\left(\frac{-t}{s}\right)\, ,
\eeq
and
\beq
c_1^{\mu}=-\left[C_F\ln\left(\frac{-t}{s}\right)
+C_A\ln\left(\frac{-u}{s}\right)+\frac{3}{4}C_F+\frac{\beta_0}{4}\right]
\ln\left(\frac{\mu_F^2}{s}\right)
+\frac{\beta_0}{2}\ln\left(\frac{\mu_R^2}{s}\right) \, .
\eeq
The expression for $\sigma^B$ can be found in Appendix D of Ref. \cite{KO2}.
The expression for $A^c$ can be easily derived from Eq. (\ref{Ac})
or by comparing
Eq. (\ref{jetqq}) with the expression in Appendix D of \cite{KO2}
for the NLO corrections for this subprocess.  

\subsubsection{$g g \rightarrow  g g$}

The NLO soft and virtual corrections are given by Eq. (\ref{jetqq})
with $c_3=2C_A$,
\beq
c_2=-2C_A\ln\left(\frac{\mu_F^2}{s}\right)
-\frac{\beta_0}{2}-4C_A \, ,
\eeq
and
\beq
c_1^{\mu}=-C_A\ln\left(\frac{p_T^2}{s}\right)
\ln\left(\frac{\mu_F^2}{s}\right)
+\frac{\beta_0}{2}\ln\left(\frac{\mu_R^2}{\mu_F^2}\right) \, .
\eeq
The expression for $\sigma^B$ can be found in Appendix E of Ref. \cite{KO2}.
The expression for $A^c$ can be easily derived from Eq. (\ref{Ac})
or by comparing
Eq. (\ref{jetqq}) with the expression in Appendix E of \cite{KO2}
for the NLO  corrections for this subprocess.  

The term $F$ in Eq. (\ref{Fterm}) can be easily derived 
for each subprocess using the matrices $H^{(0)}$, $S^{(0)}$, and
$\Gamma_S^{(1)}$ as given in Ref. \cite{KO2}.
For all subprocesses the NNLO soft and virtual corrections are given
by our master formula for complex color flows and are in agreement
with the NNLO-NLL expressions in \cite{KO2}.

\subsection{Squark and gluino production}

Our last application is the production of squarks and gluinos in hadron
colliders.
The complete NLO corrections for squark and gluino production 
in the Minimal Supersymmetric Standard Model have been given in \cite{BHSZ}.
There are several partonic subprocesses involved.

We begin with squark production. For the channel $q{\bar q} \rightarrow 
{\tilde q} {\tilde {\bar q}}$ the NLO terms $c_3$, $c_2$, and $c_1^{\mu}$ 
in the $\overline{\rm MS}$ scheme are the same as for 
the $q{\bar q} \rightarrow Q {\bar Q}$ channel in heavy quark production
in Section 4.2.1 in either 1PI or PIM kinematics, with $M=m$ the mass of
the squark. The same holds for the channel $qq \rightarrow 
{\tilde q} {\tilde q}$, and analogous results hold in the DIS scheme.
For the channel $gg \rightarrow {\tilde q} {\tilde {\bar q}}$ in the 
$\overline{\rm MS}$ scheme, $c_3$, $c_2$, and $c_1^{\mu}$ are the same as for 
the $gg \rightarrow Q {\bar Q}$ channel in heavy quark production
in Section 4.2.2 in either kinematics. 

We continue with gluino production. Here we choose $M=m$, with $m$ the mass of
the gluino. For the process
$q{\bar q} \rightarrow {\tilde g} {\tilde g}$,  $c_3$, $c_2$, and $c_1^{\mu}$
are the same as for the process 
$q{\bar q} \rightarrow {\tilde q} {\tilde {\bar q}}$ 
in either 1PI or PIM kinematics and either factorization scheme;
for the process $gg \rightarrow {\tilde g} {\tilde g}$ in the 
$\overline{\rm MS}$ scheme they are the same as for $gg \rightarrow 
{\tilde q} {\tilde {\bar q}}$ in either kinematics.

We finally have the process of squark-gluino production,
$qg \rightarrow {\tilde q} {\tilde g}$. We choose $M=m$, with $m$
the mass of the squark or the mass of the gluino. In 1PI kinematics 
in the $\overline{\rm MS}$ scheme we have 
$c_3=2(C_F+C_A)$,
\beq
c_2=-C_F-C_A-2C_F \ln\left(\frac{-u_1}{m^2}\right)
-2C_A \ln\left(\frac{-t_1}{m^2}\right)
-(C_F+C_A)\ln\left(\frac{\mu_F^2}{s}\right) \, ,
\eeq
and
\beq
c_1^{\mu}=\left[C_F\ln\left(\frac{-u_1}{m^2}\right)
+C_A \ln\left(\frac{-t_1}{m^2}\right)-\frac{3}{4}C_F-\frac{\beta_0}{4}\right]
\ln\left(\frac{\mu_F^2}{s}\right)
+\frac{\beta_0}{2}\ln\left(\frac{\mu_R^2}{s}\right)\, .
\eeq
In PIM kinematics the expressions for $c_2$ and $c_1^{\mu}$
are obtained after striking out the $\ln(-t_1/m^2)$
and  $\ln(-u_1/m^2)$ terms from the 1PI $c_2$ and $c_1^{\mu}$.

For all these processes, $A^c$, $T_1$, $T_1^c$ 
can be read off the full NLO calculation.
Also $A^c$ and $F$ can be calculated explicitly once the lowest-order
$H$, $S$, and $\Gamma_S$ matrices have been constructed.
The NNLO soft and virtual corrections are then given by 
our master formula.

\mysection{Conclusions and outlook to higher orders}

In this paper, I presented a unified approach to calculating the
NNLO soft and virtual QCD corrections for any process in hadron-hadron
and lepton-hadron collisions in either the $\overline{\rm MS}$
or DIS schemes and in either 1PI or PIM kinematics. 
The master formulas given in the paper are based on a unified threshold
resummation formalism and they allow explicit calculations for
any process, with either simple or complex color flows,
keeping in general the factorization and renormalization scales
separate and the beta function and color factors explicit. 
I verified that the scale-dependence of 
the physical cross section cancels out at NNLO for any process.
Detailed results, illustrating the use of the master formulas, 
were given for various electroweak, Higgs, QCD, and SUSY processes in 
various factorization schemes, kinematics, and colliders. 
As tests of the master formulas, I reproduced the previously known
NNLO corrections for Drell-Yan and Higgs production, deep inelastic
scattering, and $W^+ \gamma$ production, thus also determining a number 
of two-loop anomalous dimensions and other quantities which are
needed in NNLL resummations. 
Furthermore, I presented new results for several other processes
in the Standard Model and beyond. 

The NNLO corrections increase theoretical accuracy and diminish the dependence
on the factorization and renormalization scales, and thus are essential 
in further testing QCD and particularly
in searching for the Higgs boson and supersymmetric particles
as well as other processes, such as flavor-changing neutral currents,
that signal new physics beyond the Standard Model.

The unified approach employed in this paper can be extended to higher
orders. For a sketch of how this extension may be carried out through
next-to-next-to-next-to-next-to-leading order in the
specific context of heavy quark hadroproduction see Ref. \cite{NK}.
A complete calculation of next-to-next-to-next-to-leading and even 
higher-order soft and virtual corrections is a formidable task and unlikely 
to be necessary at least in the forseeable future. The accuracy attainable
at NNLO should be sufficient for the high-energy colliders of our era.
The unified master formulas for the NNLO soft and virtual corrections
for any process, which are an important component of the full NNLO 
calculation, should serve as a milestone in the push for ever-increasing 
theoretical accuracy and understanding of high-energy processes.
  
\mysection*{Acknowledgements}

The author's research has been supported by a Marie Curie Fellowship of 
the European Community programme ``Improving Human Research Potential'' 
under contract number HPMF-CT-2001-01221.

\end{document}